# Modeling water relative permeability in uniform grain packs

Behzad Ghanbarian

Porous Media Research Lab, Department of Geology, Kansas State University, Manhattan KS 66506 USA

Email address: ghanbarian@ksu.edu

**Abstract**

Accurate estimation of water relative permeability has been of great interest in various research areas because of its broad applications in soil physics and hydrology as well as oil and gas production and recovery. Critical path analysis (CPA), a promising technique from statistical physics, is well known to be applicable to heterogeneous media with broad conductance or pore size distribution (PSD). By heterogeneity, we mean variations in the geometrical properties of pore space. In this study, we demonstrate that CPA is also applicable to packings of spheres of the same size, known as homogeneous porous media. More specifically, we apply CPA to model water relative permeability ($k_{rw}$) in mono-sized sphere packs whose PSDs are fairly narrow. We estimate the $k_{rw}$ from (1) the PSD and (2) the PSD and saturation-dependent electrical conductivity ($\sigma_r$) for both drainage and imbibition processes. We show that the PSD of mono-sized sphere packs approximately follows the log-normal probability density function. Comparison with numerical simulations indicate that both the imbibition and drainage $k_{rw}$ are estimated from the PSD and $\sigma_r$ data more accurately than those from the PSD. We show that CPA can estimate $k_{rw}$ in mono-sized sphere packs precisely.

**1. Introduction**

Water relative permeability, $k_{rw}$, and its concept appear in the simultaneous flow of two immiscible fluids in porous media as well as oil production and recovery in rock reservoirs. Therefore, accurate modeling and estimation of relative permeability of water in water-oil or water-gas systems have been of great interest in petroleum and chemical engineering as well as hydrology and soil physics.

In order to better understand the concepts of two-phase flow in natural rocks and sediments with complex pore structure, less disordered porous materials e.g., Finney's and other types of sphere packings as well as glass beads have been extensively investigated in the literature (see e.g., Naar et al., 1962; Topp and Miller, 1966; Bryant and Blunt, 1992; Pan et al., 2004; Silin and Patzek, 2009; Dye et al., 2016). However, sphere packs are different from real rocks or soils in several ways. For example, in sphere packs grain- and pore-size distributions are much narrower than those in fracture networks and rocks. Grains are quite spherical, and pores are typically sinusoidal in shape (see e.g., Hopkins and Ng (1986) and Feng et al. (1987)). In addition, the pore-solid interface of sphere packs is far smoother compared to pore wall roughness in rocks, and thus specific surface area may be substantially different. Nonetheless, mono-sized and bi-disperse sphere packs have been commonly used to study relative permeability and other transport phenomena in the literature. For example, in order to demonstrate that consolidated and unconsolidated porous media show different imbibition flow behavior, Naar et al. (1962) measured oil and gas relative permeabilities in a binary mixture of spheres (with relatively wide pore-size distribution) and a poorly sorted consolidated sandstone. Their results exhibited that at a given saturation the relative permeability of the non-wetting phase during imbibition was less than that during drainage for a consolidated medium, while the opposite was true for a unconsolidated medium. Naar et al. (1962) found similar results for the wetting phase and argued that the difference in imbibition behavior could be related to pore size distribution and cementation.

More recently, Li et al. (2005), Silin and Patzek (2009), Landry et al. (2014), and Mawer et al. (2015) addressed relative permeability in sphere packs by means of numerical simulations. Li et al. (2005) simulated two-phase flow in a homogeneous sphere pack with porosity of 0.36 and the relative standards deviation of the spherical grain size of near 10%. They used a multiple-relaxation-time Shan-Chen type multicomponent LB model and reported correlations of the relative permeabilities as a function of capillary number, wettability, and fluid viscosity.

For decades, the simultaneous flow of two immiscible phases through porous media lacked a sound theoretical description due to the complex geometry and topology of the pore space (Larson et al., 1981). However, percolation theory provided a powerful theoretic approach to investigate interconnectivity and its effects on flow and transport in networks and porous media (Sahimi, 1994; 2011; Hunt et al., 2014). Percolation theory was successfully used to compute relative permeability in random porous materials (see e.g., Larson et al., 1981;

Chatzis and Dullien, 1985; Jerauld and Salter, 1990; Heiba et al., 1982; 1992; Blunt et al., 1992; Hunt, 2001; Sahimi, 1994, 2011; Hunt et al., 2014; Ghanbarian et al., 2015a). For a recent review, see Hunt and Sahimi (2017).

Larson et al. (1981) were probably first who modeled various properties of flow and transport e.g., intrinsic and relative permeabilities, residual saturation, etc. using percolation theory. Heiba et al. (1982; 1992) applied network models and percolation theory to model two-phase relative permeability in porous media. They showed that their percolation-based model could predict the typical trends of two-phase relative permeability in random porous media satisfactorily.

Concepts from critical path analysis (CPA) and percolation theory have been successfully applied to estimate water relative permeability in heterogeneous porous media e.g., as soils and rocks with broad pore size distribution (Hunt, 2001; Ghanbarian-Alavijeh and Hunt 2012a; Ghanbarian and Hunt, 2017; Ghanbarian et al., 2016; 2017b). Hunt (2001) was first to address water relative permeability modeling by means of CPA. His approach, generalized later by Ghanbarian-Alavijeh and Hunt (2012a) and Hunt et al. (2013), is based on the fact that flow and transport in porous media with broad distribution of conductances (or equivalently pore sizes) is dominated by those conductances with magnitudes greater than some critical value. Hunt (2001) estimated water relative permeability from capillary pressure curve and showed good match with experiments from the Hanford site, particularly at high water saturations. A comprehensive review of CPA-based water relative permeability models can be found in Ghanbarian et al. (2015a) and Ghanbarian and Hunt (2017).

To the best of the author's knowledge, it is not clear whether CPA can accurately estimate water relative permeability in mono-sized sphere packs with relatively narrow pore size distribution. Therefore, the main objectives of this study are: (1) using CPA to estimate water relative permeability $k_{rw}$ in packings of *randomly distributed* spheres of the same size, and (2) comparing $k_{rw}$ estimated from the PSD with that predicted from the PSD and $\sigma_r$. We also investigate hysteresis in the $k_{rw}$-$S_w$ relation in packings of mono-sized spheres. By comparison with numerical simulations, we demonstrate that CPA estimates $k_{rw}$ in homogeneous porous media accurately.

## 2. Theory

In this section, we first present the log-normal probability density function to describe the pore size distribution in mono-sized sphere packs. We then apply concepts from critical path analysis to model water relative permeability.

### 2.1. Pore size distribution $f(r)$

To derive analytical relationships for capillary pressure curve and water relative permeability, one needs to represent the pore size distribution (PSD) using a probability density function. For this purpose, we use the log-normal probability density function that has been frequently applied to porous media and fracture networks (see e.g., Kosugi 1994; 1996; Madadi and Sahimi, 2003):

$$f(r) = \frac{A}{\sqrt{2\pi}\sigma r} \exp\left[-\left(\frac{\ln\left(\frac{r}{r_m}\right)}{\sqrt{2}\sigma}\right)^2\right], \quad r_{\min} \leq r \leq r_{\max}. \tag{1}$$

In Eq. (1), $r$ is the pore radius, $r_{\min}$ and $r_{\max}$ are the smallest and largest pore radii, respectively, representing the lower and upper bounds of the log-normal distribution, $r_m$ is the geometric mean pore radius, $\sigma$ is the log-normal standard deviation, and $A$ is a normalizing prefactor.

One may assume that each pore is occupied by either water or air. Accordingly, in water-wet systems, water occupies the smallest pores, while air fills the largest pores when all the pores are simultaneously accessible. However, such a pore occupancy hypothesis is only an approximation because there are normally thin films of water residing in pores that are occupied by the non-wetting fluid (in this case, air).

If one follows Eq. (1) and then invokes the Young-Laplace equation, the capillary pressure curve model corresponding to the log-normal distribution is (Ghanbarian et al., 2018b)

$$S_\text{e} = \frac{\text{erf}\left(\frac{3\sigma^2 - \ln\left(\frac{P_\text{m}}{P_\text{max}}\right)}{\sqrt{2}\sigma}\right) - \text{erf}\left(\frac{3\sigma^2 - \ln\left(\frac{P_\text{m}}{P}\right)}{\sqrt{2}\sigma}\right)}{\text{erf}\left(\frac{3\sigma^2 - \ln\left(\frac{P_\text{m}}{P_\text{max}}\right)}{\sqrt{2}\sigma}\right) - \text{erf}\left(\frac{3\sigma^2 - \ln\left(\frac{P_\text{m}}{P_\text{min}}\right)}{\sqrt{2}\sigma}\right)}, \quad P_\text{min} \leq P \leq P_\text{max} \tag{2}$$

where the effective water saturation $S_\text{e} = (S_\text{w} - S_\text{wr})/(1 - S_\text{wr})$ in which $S_\text{w}$ is the water saturation and $S_\text{wr}$ is the residual water saturation, erf is the error function, $P_\text{m}$ is the capillary pressure corresponding to the geometric mean pore radius ($r_\text{m}$), and $P_\text{min}$ and $P_\text{max}$ are the minimum and maximum capillary pressures, respectively.

**2.2. Critical path analysis (CPA)**

Ambegaokar et al. (1971) argued that transport in a disordered system with a broad distribution of conductances, $f(g)$, is controlled by those with magnitudes greater than some critical conductance $g_\text{c}$, defined as the smallest conductance required to form a conducting sample-spanning (or infinite connected) cluster. Within the CPA framework, transport in a network of pores is dominated by highly-conducting pores, while low-conducting ones have trivial contribution to the overall transport (Hunt et al., 2014). Imagine a pore network constructed of pores of various sizes shown in Fig. 2a. Following Friedman and Seaton (1998), let us remove all pores from the network. We then replace them in their original locations in a decreasing order from the largest to the smallest pore size. As the first largest pores are replaced, there is still no percolating cluster (Fig. 2b). However, after a sufficiently large fraction of pores are replaced within the network, a sample-spanning cluster forms and the system starts percolating (Fig. 2c).

CPA can only be used to model the wetting-phase relative permeability in porous media because the wetting and non-wetting phases respectively occupy the smallest and largest pores. The smallest pores are normally ignored by the CPA, while the largest ones are important to CPA. For non-wetting relative permeability, however, one may apply universal quadradic scaling from percolation theory (Hunt, 2005; Ghanbarian-Alavijeh and Hunt, 2012b), a combination of percolation theory and effective-medium approximation (Ghanbarian et al., 2018a), or the effective-medium approximation (Ghanbarian, 2018). For a fully-saturated medium, the minimum water saturation required to form the sample spanning cluster and let the system percolates has two contributors: (1) the water saturation needed to fill pores between $r_c$ and $r_{max}$, and (2) the residual water saturation, $S_{wr}$ (Heiba et al., 1982; 1992; Ghanbarian et al., 2017b). By applying concepts from CPA and assuming that PSD follows the log-normal probability density function, Eq. (1), one may define the critical water saturation for water permeability, $S_{wc}$, under fully-saturated conditions as follows:

$$S_{wc} = \frac{1}{\phi} \int_{r_c(S_w=1)}^{r_{max}} sr^3 f(r) dr + S_{wr} \qquad (3)$$

$$= \frac{sAr_m^3}{2\phi} \exp\left(\frac{9\sigma^2}{2}\right) \left[\text{erf}\left(\frac{3\sigma^2 - \ln\left(\frac{r_c(S_w=1)}{r_m}\right)}{\sqrt{2}\sigma}\right) - \text{erf}\left(\frac{3\sigma^2 - \ln\left(\frac{r_{max}}{r_m}\right)}{\sqrt{2}\sigma}\right)\right] + S_{wr}$$

where $s$ is the shape factor and $\phi$ is the porosity.

Under partially-saturated conditions one has

$$S_{wc} = \frac{1}{\phi} \int_{r_c(S_w)}^{r} sr^3 f(r) dr + S_{wr} \tag{4}$$

$$= \frac{sAr_m^3}{2\phi} \exp\left(\frac{9\sigma^2}{2}\right) \left[\text{erf}\left(\frac{3\sigma^2 - \ln\left(\frac{r_c(S_w)}{r_m}\right)}{\sqrt{2}\sigma}\right) - \text{erf}\left(\frac{3\sigma^2 - \ln\left(\frac{r}{r_m}\right)}{\sqrt{2}\sigma}\right)\right] + S_{wr}$$

Arranging Eqs. (3) and (4) in terms of $r_c(S_w = 1)$ and $r_c(S_w)$ yields

$$r_c(S_w = 1) = r_m \exp\left[3\sigma^2 - \sqrt{2\sigma^2}\,\text{erf}^{-1}\left(\frac{\phi(S_{wc} - S_{wr})}{C} + \text{erf}\left(\frac{3\sigma^2 - \ln\left(\frac{r_{max}}{r_m}\right)}{\sqrt{2}\sigma}\right)\right)\right] \tag{5}$$

$$r_c(S_w) = r_m \exp\left[3\sigma^2 - \sqrt{2\sigma^2}\,\text{erf}^{-1}\left(\frac{\phi(S_{wc} - S_{wr})}{C} + \text{erf}\left(\frac{3\sigma^2 - \ln\left(\frac{r}{r_m}\right)}{\sqrt{2}\sigma}\right)\right)\right] \tag{6}$$

in which $C = \frac{sAr_m^3}{2} \exp\left(\frac{9\sigma^2}{2}\right)$.

In what follows, we propose to estimate water relative permeability from: (1) the PSD, and (2) the PSD and electrical conductivity ($\sigma_r$) data.

### 2.3. Estimating water relative permeability from pore size distribution

Within the critical path analysis framework, the water permeability $k_w$ is proportional to the critical hydraulic conductance i.e., $k_w \propto g_c$ (Hunt, 2001; Ghanbarian-Alavijeh and Hunt, 2012a). The latter, $g_c$, is related to the critical pore radius via the Poiseuille equation ($g_c \propto r_c^\alpha$). For a perfectly cylindrical pore whose length is uncorrelated to its radius, $\alpha = 4$. Accordingly, one may define water relative permeability $k_{rw}$ as follows:

$$k_{rw} = \frac{k_w(S_w)}{k_w(S_w=1)} = \frac{g_c(S_w)}{g_c(S_w=1)} = \left(\frac{r_c(S_w)}{r_c(S_w=1)}\right)^\alpha \tag{7a}$$

where $k_w(S_w)$ and $k_w(S_w = 1)$ are water permeabilities under partially and fully saturated conditions, respectively. If $r_c(S_w = 1)$ and $r_c(S_w)$ in Eq. (7a) are respectively replaced with those given in Eqs. (5) and (6), one has (Ghanbarian et al., 2018b)

$$k_{rw} = \left[\frac{\exp\left(3\sigma^2 - \sqrt{2\sigma^2}\,\mathrm{erf}^{-1}\left(\frac{\phi(S_{wc}-S_{wr})}{C} + \mathrm{erf}\left(\frac{3\sigma^2 - \ln\left(\frac{r}{r_m}\right)}{\sqrt{2}\sigma}\right)\right)\right)}{\exp\left(3\sigma^2 - \sqrt{2\sigma^2}\,\mathrm{erf}^{-1}\left(\frac{\phi(S_{wc}-S_{wr})}{C} + \mathrm{erf}\left(\frac{3\sigma^2 - \ln\left(\frac{r_{max}}{r_m}\right)}{\sqrt{2}\sigma}\right)\right)\right)}\right]^{\alpha} \quad (7b)$$

in which $r$ is related to $S_w$ via Eq. (2). Recall that $\alpha$ is the exponent in the Poiseuille equation. Recently, Ghanbarian et al. (2016) proposed a theoretical scaling of Poiseuille's law for flow in cylindrical pores with irregular rough surfaces. They showed that $\alpha$ should be equal to 3 in media with smooth pore-solid interface (see their Eq. (14)). To estimate $k_{rw}$ from the PSD via Eq. (7), we set $\alpha = 3$. This assumption is consistent with single-phase permeability estimations in uniform grain packs (Ghanbarian, 2019).

**2.4. Estimating water relative permeability from pore size distribution and electrical conductivity**

Similarities between hydraulic and electrical flow in porous media directed attentions to estimating water relative permeability from electrical conductivity data, which are easily and routinely measured through petrophysical evaluations. For instance, Rose and Bruce (1949) extended the Kozeny-Carman model to unsaturated media and linked $k_{rw}$ to electrical resistivity measurements. Later, Wyllie and Spangler (1952) revisited the Rose and Bruce (1949) model and showed that their revised model estimated $k_{rw}$ from capillary pressure data and electrical resistivity measurements accurately in unconsolidated sands. Katz and Thompson (1986), followed by Friedman and Seaton (1998), Hunt (2001),

Ghanbarian et al. (2016;2017a) and many others, showed that water permeability, $k_w$, is proportional to the product of bulk electrical conductivity and critical pore radius squared under fully saturated conditions ($k_w(S_w = 1) \propto \sigma_b(S_w = 1)r_c^2(S_w = 1)$). Expecting a similar relationship to hold for the water permeability at partial saturation (see e.g., Doussan and Ruy (2009) and Ghanbarian et al. (2017b)), one has

$$k_{rw} = \frac{k_w(S_w)}{k_w(S_w=1)} = \frac{\sigma_b(S_w)}{\sigma_b(S_w=1)} \left(\frac{r_c(S_w)}{r_c(S_w=1)}\right)^2 \tag{8a}$$

where $\sigma_b(S_w)$ is the bulk electrical conductivity under partially saturated conditions and $\sigma_r = (\sigma_b(S_w)/\sigma_b(S_w = 1))$ is the relative electrical conductivity. Substituting $r_c(S_w = 1)$ and $r_c(S_w)$ from Eqs. (5) and (6) into Eq. (8a) gives

$$k_{rw} = \sigma_r \left[\frac{\exp\left(3\sigma^2 - \sqrt{2\sigma^2}\,\mathrm{erf}^{-1}\left(\frac{\phi(S_{wc}-S_{wr})}{C} + \mathrm{erf}\left(\frac{3\sigma^2 - \ln\left(\frac{r}{r_m}\right)}{\sqrt{2}\sigma}\right)\right)\right)}{\exp\left(3\sigma^2 - \sqrt{2\sigma^2}\,\mathrm{erf}^{-1}\left(\frac{\phi(S_{wc}-S_{wr})}{C} + \mathrm{erf}\left(\frac{3\sigma^2 - \ln\left(\frac{r_{max}}{r_m}\right)}{\sqrt{2}\sigma}\right)\right)\right)}\right]^2 \tag{8b}$$

Eq. (8b) holds only for porous media in which the electric and hydraulic percolation sets are identical (Le Doussal, 1989) and the percolation threshold for the two transport mechanisms is the same (Katz and Thompson, 1986). For example, when surface conduction exists some of the electric current flows through electric double layers carrying excess ionic charges. In such a case, the percolation threshold for electrical conductivity would be smaller than that for hydraulic conductivity because pore surfaces contribute effectively to electric current but not to water flow particularly at low water saturations. Likewise, in porous media with significant film and corner flow, one should expect the two percolation thresholds to be different.

We should point out that Eq. (7b) was earlier developed by Ghanbarian et al. (2018b) and successfully evaluated to estimate imbibition water relative permeability from pore size distribution derived from images in coated and uncoated papers. However, Eq. (8b) has not been proposed and/or assessed with experiments or simulations before. In addition, $k_{rw}$ estimations by Eqs. (7b) and (8b) have not been previously compared.

**2.5. Universal power-law scaling from percolation theory**

CPA describes the saturation dependence of the minimum hydraulic conductance on the most conductive flow paths ($g > g_c$). As water saturation reduces toward $S_{wc}$, however, the most important limitation to the water permeability is no longer the rate-limiting conductance of the most conductive paths (Hunt and Gee, 2003). Near and above $S_{wc}$, however, results from topology and percolation scaling can be applied to characterize $k_{rw}$ and its tendency to 0. Therefore, there exists a crossover water saturation ($S_{wx}$) above which $k_{rw}$ is dominated by a rate-limiting conductance (CPA). Below $S_{wx}$, however, the water relative permeability is determined by the following universal power-law scaling

$$k_{rw} = k_0 (S_w - S_{wc})^t, \qquad S_{wc} < S_w \leq S_{wx} \tag{9}$$

where $k_0$ is a constant coefficient and the exponent $t$ is the universal scaling exponent for water percolation. The value of $t$ only depends on the dimensionality of the system; $t = 1.3$ in two and $t = 2$ in three dimensions (Stauffer and Aharony, 1994). We should point out that an exponent different than 1.3 or 2 in Eq. (9) refers to non-universal scaling from percolation theory (Feng et al., 1987).

We use the CPA scaling, Eqs. (7b) and (8b), for estimating $k_{rw}$ at high to intermediate water saturations, and apply the universal power-law scaling, Eq. (9), to determine $k_{rw}$ at low $S_w$ values near the critical water saturation $S_{wc}$. Accordingly, one should expect a crossover between the CPA scaling and the universal power-law scaling to occur at some water saturation $S_{wx}$. The necessity of crossing over from the CPA regime to the universal power-law scaling was first invoked by Hunt and Gee (2003).

In the following, we first describe the numerical simulations in sphere packs, compare the CPA-based models – Eqs. (7b) and (8b) in combination with Eq. (9) – with the simulations, and then discuss the obtained results.

**3. Numerical simulations**

The data used in this study are numerical simulations in mono-sized sphere packs from Mawer et al. (2015) and Silin and Patzek (2009). We briefly describe each dataset, and the interested reader is referred to the original published articles for further details.

**- Numerical simulations of Mawer et al. (2015)**

Mawer et al. (2015) performed numerical simulation of saturation-dependent capillary pressure, electrical conductivity and water permeability for the Finney pack (Finney, 1970) with $\phi = 0.362$, and 14 other packs whose porosities ranged between 0.23 and 0.46 (see Table 1).

To generate a partially-saturated pack, Mawer et al. (2015) saturated the pore space and carried out drainage and imbibition simulations to capture hysteresis. The electrical conductivity of each pack was computed using the finite-element approach (Garboczi, 1998). Water permeability for each partially-saturated pack was determined using the lattice-Boltzmann method. Previous comparisons with experiment have shown that the lattice-Boltzmann method can accurately simulate water relative permeability in unconsolidated (Li et al., 2005; Hao and Cheng, 2010) and consolidated (Ramstad et al., 2012; Shikhov et al., 2017) porous media. Further details are given by Mawer et al. (2015). To estimate the water relative permeability, the imbibition/drainage capillary pressure curves were first converted to the pore size distributions. Although it is relatively straightforward to measure and/or simulate capillary pressure curve, inferring pore size distribution from that is not simple. The pore size distribution is the probability density function that yields the distribution of pore volume by an effective or characteristics pore size. In this study, $\Delta S_w/\Delta \ln(P)$ was plotted versus pore radius, and parameters of the log-normal distribution, namely, $A$, $r_m$, and $\sigma$, were determined by fitting Eq. (1) to the derived pore size distributions. $r_{min}$ was set equal to zero for all packs, while $r_{max}$ was determined from the fitted log-normal distribution to the derived PSD and more specifically from where the log-normal distribution touches the $x$ axis. We should point out that the proposed $k_{rw}$ models (Eqs. 7b and 8b) and their estimations are not greatly sensitive to these two parameters.

Salient properties of the Finney pack and 14 other sphere packs are given in Table 1. The residual water saturation was estimated from the dry end of the simulated capillary pressure curves and set equal to 0 ($S_{wr}$ = 0). Within continuum percolation theory, percolation threshold of a randomly distributed pack of spheres is 0.03 (van der Marck, 1996; Rintoul, 2000). We, accordingly, set $S_{wc}$ = 0.03 for all packs from Mawer et al. (2015).

$k_0$ in Eq. (9) was determined by setting Eqs. (9) and (7b) or (8b) equal at some crossover water saturation $S_{wx}$. The value of $S_{wx}$ can be calculated by setting the first derivative of the universal power-law equation equal to that of either Eq. (7b) or (8b). In this study, the values of $k_0$ and $S_{wx}$ were numerically computed. For this purpose, we first fitted the spline function to the $k_{rw}$ values estimated via Eq. (7b) or (8b). We then calculated $k_0$ and determined the slope at each water saturation numerically. The crossover point is the water saturation at which the slope of Eq. (7b) or (8b) is equal to that of Eq. (9). Accordingly, to estimate the water relative permeability over the entire range of $S_w$, we used either Eq. (7b) or (8b) for $S_{wx} \leq S_w \leq 1$, and Eq. (9) for $S_{wc} \leq S_w \leq S_{wx}$.

**- Numerical simulations of Silin and Patzek (2009)**

Silin and Patzek (2009) applied the finite difference method to simulate the drainage water relative permeability in the Finney pack under water-wet conditions. They assumed that capillary pressure determines fluid distribution, and numerical simulation of the Stokes equations evaluates the pore-scale flow field. To reduce the computational intensity, the medium was partitioned into layers, and the harmonic mean of permeabilities computed for various layers was used to determine the permeability of the entire medium.

The capillary pressure curve and the saturation-dependent electrical conductivity are not available for the Finney pack from Silin and Patzek (2009). Therefore, to estimate $k_{rw}$ we used the log-normal PSD parameters, reported for the Finney pack given in Table 1, and the simulated $\sigma_r$ from Mawer et al. (2015). In another study, Ghanbarian and Sahimi (2017) analyzed the saturation-dependent electrical conductivity curves from Mawer et al. (2015). They showed that although simulated in various mono-sized sphere packs, all the drainage electrical conductivity data collapsed into a single curve and followed the universal power-law scaling from percolation theory, similar to Eq. (9) with $t = 2$ and $S_{wc} = 0$ (see their Fig. 1). This clearly indicates that electrical flow in uniform packs is strongly controlled by topological characteristics such as pore connectivity as well as tortuosity.

## 4. Results

In this section, we present the results obtained from comparing our theoretical models – Eqs. (7b) and (8b) in combination with Eq. (9) – with numerical simulations and discuss the water relative permeability modeling and estimation via CPA in sphere packs.

### 4.1. $k_{rw}$ estimations for the Mawer et al. (2015) data

#### 4.1.1. Water relative permeability during drainage

Figure 2 shows the PSDs derived from the simulated capillary pressure curves under drainage conditions. As can be seen, the log-normal probability density function approximately represents the PSD accurately ($R_2 > 0.93$). The average standard deviation value of the log-normal probability density function is 0.206 ($0.156 \leq \sigma \leq 0.279$; see Table 1). Comparing $\sigma = 0.206$ with $\sigma = 2.87$ reported by Kosugi (1996) for the Beit Netofa clay soil (see his Table 1) indicates that the mono-sized sphere packs studied here have relatively narrow PSDs. $\sigma = 0.206$ is also remarkably less than $\sigma = 1.084$, 2.110, 2.333, 2.395, and 2.916 reported by Hwang and Choi (2006) respectively for coarse, moderately coarse, medium, moderately fine, and fine soil texture classes (see their Table 2).

Figure 3 presents the $k_{rw}$ curves estimated from the PSD, represented by black and gray lines, as well as the PSD and $\sigma_r$, denoted by blue and red lines. The RMSLE values reported in each plot is the root mean square logarithmic error. For all sphere packs, $k_{rw}$ was estimated from the PSD and $\sigma_r$ data more accurately than that estimated from the PSD. Although both approaches estimated $k_{rw}$ accurately and similarly at high water saturations ($S_w \gtrsim 0.8$), using the PSD and $\sigma_r$ data we found more accurate estimations at intermediate and low $S_w$ values. As Fig. 3 indicates, applying the PSD resulted in the water relative permeability overestimation for $S_w \lesssim 0.8$.

Electrical flow has been previously linked to hydraulic flow in porous media (Katz and Thompson, 1986; Friedman and Seaton, 1998) because it is routinely measured in reservoir characterization. Our results, in accord with those from the literature (Wyllie and Spangler, 1952; Doussan and Ruy, 2009; Ghanbarian et al., 2017b), indicate that incorporating electrical conductivity in addition to capillary pressure curve improve relative permeability estimations mainly because electrical flow is influenced by pore space geometrical and topological properties affecting hydraulic flow.

Conceptually, CPA is an appropriate method for upscaling flow and transport in porous media with broad conductance or pore size distributions (Katz and Thompson, 1986). However, the term "broad" has not been satisfactorily defined in the literature. Regarding the validity of the critical path analysis approach, Shah and Yortsos (1996) stated that, "The basic argument underlying this theory [i.e., critical path analysis] is that because of the large exponent in the pore conductance-pore radius relationship, $g \sim r^4$, natural porous media, even though moderately disordered in pore size, possess a wide conductance distribution." For example, in a relatively homogenous medium whose maximum pore size is only ten times greater than its minimum pore size, the ratio $g_{max}/g_{min} = 10^4$, corresponding to a relatively broad conductance distribution, $f(g)$. Accordingly, even for mono-sized sphere packs, one may expect critical path analysis to be reasonably accurate.

Figure 3 also indicates that the crossover water saturations for the two approaches used to estimate $k_{rw}$ are remarkably different. When $k_{rw}$ was estimated from only the PSD (Eq. 7b), we found $S_{wx} \simeq 0.8$, while $S_{wx}$ for the $k_{rw}$ estimation from the PSD and $\sigma_r$ (Eq. 8b) was near 0.12. Differences between Eqs. (7b) and (8b) are in their: (1) exponents and (2) prefactors. The exponent $\alpha = 3$ in Eq. (7b) is greater than 2 in Eq. (8b), and thus $k_{rw}$ estimates via Eq. (7b) should be less than that by Eq. (8b) at intermediate to high saturations. However, the effect of prefactor is evidently more profound (see Figs. 3 and 6). The prefactor is 1 in Eq. (7b), while equal to the relative electrical conductivity ($\sigma_r$) in Eq. (8b). This is probably why the crossover water saturations are remarkably different.

In Fig. 4, we show the capillary pressure curves, pore size distributions, and water relative permeability curves for the Finney pack and 14 sphere packs from Mawer et al. (2015). Although the capillary pressure curves (Fig. 4a) and pore size distributions (Fig. 4b) are relatively scattered, the simulated $k_{rw}$ data for all the packs collapsed into a single curve (see Figs. 4c and 4d).

If the Kozeny-Carman model $k_w(S_w = 1) \propto \phi^3$, known to be valid in mono-sized packs, is adopted for partially-saturated media, one has $k_{rw} = S_w^3$. Similar power laws with slightly different exponents e.g., 3.5 (Averjanov, 1950) and 4 (Corey, 1954) were previously used to describe the saturation dependence of wetting-phase permeability in consolidated porous media (see also Brutsaert, 1967; Mualem, 1976).

The black line in Figs. 4c and 4d represents the adopted Kozeny-Carman equation ($k_{rw} = S_w^3$), also known as the cubic model. Although $k_{rw} = S_w^3$ estimated the drainage water relative permeability accurately at high water saturations, it overestimated $k_{rw}$ at intermediate and low $S_w$ values. Hao and Cheng (2010) found that at higher saturations ($S_w \gtrsim 0.5$) the cubic model underestimated the water relative permeability simulated via the lattice-Boltzmann method in a sphere pack, while it overestimated $k_{rw}$ for $S_w \lesssim 0.5$ (see their Fig. 1b).

From Fig. 4d, it is clear that the value of critical water saturation should not be zero, as assumed in the adopted Kozeny-Carman model. By fitting the non-universal power-law model from continuum percolation theory (Sahimi, 2011; Hunt et al., 2014), we found that $k_{rw} = [(S_w - S_{wc})/(1 - S_{wc})]^\mu$ with $S_{wc} = 0.05$ and $\mu = 3.3$ fit the simulated water relative permeability data accurately over the entire range of water saturation (results not shown). This indicates the nontrivial effect of non-zero critical water saturation on the $k_{rw}$ estimation in mono-sized sphere packs. The non-universal exponent $\mu > 2$ clearly shows that $k_{rw}$ is not only a function of water saturation but also pore size distribution. When the effect of pore size distribution is minimal, one should expect $\mu$ to be equal to 2.

**4.1.2. Water relative permeability during imbibition**

The log-normal probability density function, Eq. (1), and the pore size distribution of the mono-sized sphere packs derived from the simulated imbibition capillary pressure curve are presented in Fig. 5. As seen, Eq. (1) fitted the PSDs relatively well ($R_2 > 0.88$; Table 1). The log-normal standard deviation value ranged between 0.315 and 0.429 ($0.315 \leq \sigma \leq 0.429$). Comparing the drainage log-normal standard deviation with the imbibition one shows that the imbibition PSDs are broader than the drainage ones. Although the average drainage log-normal standard deviation $\sigma = 0.206$ is remarkably less than the average imbibition log-normal standard deviation $\sigma = 0.369$, the latter is still substantially less than those reported for natural porous media by Kosugi (1996) and Hwang and Choi (2006).

Interestingly, Kosugi (1996) found that the imbibition log-normal standard deviation for the Guelph loam was less than the drainage one (0.649 vs. 0.966; see his Table 1). Results of the $k_{rw}$ estimations via the CPA-based models, Eqs. (7b) and (8b) in combination with Eq. (9), are given in Fig. 6. As this figure shows, similar to the drainage results, $k_{rw}$ was estimated from the PSD and $\sigma_r$ data more accurately than that estimated from the PSD (see RMSLE values reported in each plot). Although both approaches estimated the $k_{rw}$ precisely and similarly at high water saturations ($S_w \gtrsim 0.8$), $k_{rw}$ was slightly overestimated from the PSD at intermediate and low water saturations ($S_w \lesssim 0.8$).

Figure 7 shows the simulated capillary pressure curves and their corresponding PSDs as well as the simulated water relative permeability curves for the Finney pack and 14 sphere packs from Mawer et al. (2015). Similar to the drainage process (Fig. 4), although the capillary pressure curves and PSDs are relatively scattered, all the water relative permeability data collapsed into a single curve (Fig. 7c and 7d). This clearly indicates that the effect of the PSD on $k_{rw}$ is minimal in mono-sized sphere packs. The black line represents the Kozeny-Carman model adopted for partially-saturated conditions (i.e., $k_{rw} = S_w^3$). Although it estimated the $k_{rw}$ at high water saturations accurately, the adopted Kozeny-Carman model overestimated the water relative permeability at intermediate and low $S_w$ values. Figure 7d clearly shows that the value of the critical water saturation should be small but non-zero, which is due to ignoring thin water films. We found that the power-law model $k_{rw} = [(S_w - S_{wc})/(1 - S_{wc})]^\mu$ with $S_{wc} = 0.03$ and $\mu = 3.5$ fitted the imbibition water relative permeability data accurately (results not shown). Interestingly, the values $S_{wc} = 0.03$ and $\mu = 3.5$ determined for the imbibition process are not greatly different than $S_{wc} = 0.05$ and $\mu = 3.3$ calculated for the drainage process. Accordingly, the effect of hysteresis on the saturation-dependent water relative permeability curves in mono-sized sphere packs is not remarkable. This is in accord with the experimental observations of Topp and Miller (1966) who measured water relative permeability in mono-sized glass beads under unsteady-state conditions. They stated that, "Large hysteresis was recorded in the relation of pressure to saturation or to [hydraulic] conductivity. The relation of [hydraulic] conductivity to saturation showed hysteresis which, though significantly larger than the experimental error, would for most practical purposes be negligible." Similar results were obtained by Blunt (1997) who simulated relative permeability in three-dimensional pore networks with narrow pore size distributions.

**4.2. $k_{rw}$ estimations for the Silin and Patzek (2009) data**

We also compare the $k_{rw}$ estimates via our CPA-based models, Eqs. (7b) and (8b) in combination with Eq. (9), with the numerical simulations in the Finney pack from Silin and Patzek (2009). The pore size distribution and electrical conductivity of the pack in the study of Silin and Patzek (2009) are not available. Accordingly, we used parameters reported for the Finney pack in the Mawer et al. (2015) article, which are similar to those reported by Dadvar and Sahimi (2003). Figure 8 shows that $k_{rw}$ was estimated accurately from the PSD and $\sigma_r$ data. However, using only the PSD, we found that the CPA model, Eq. (7) in combination with Eq. (9), overestimated $k_{rw}$ at intermediate and low water saturations, in accord with the results obtained from the Mawer et al. (2015) database (Figs. 3 and 6).

**5. Further Discussion**

The precise estimation of water relative permeability requires including the effect of numerous factors, such as the pore geometry, wettability, accessibility and connectivity, and viscous and capillary forces. However, the quantification of integrated influences of all such factors within a theoretic approach is challenging. One should, therefore, not expect one theory e.g., CPA to estimate $k_{rw}$ in porous media under any conditions accurately. The proposed CPA-models (Eqs. 7b and 8b) relate $k_{rw}$ to water saturation ($S_w$), porosity ($\phi$), pore size distribution ($f(r)$), pore connectivity (reflected in $S_{wc}$), residual water saturation ($S_{wr}$), minimum and maximum pore sizes ($r_{min}$ and $r_{max}$), pore shape geometry ($s$), and/or relative electrical conductivity ($\sigma_r$). The effect of wettability is implicitly incorporated since the knowledge of contact angle is required to convert capillary pressure curve into pore size distribution. In the following, we further discuss various factors and their influences on critical water saturation ($S_{wc}$) and water relative permeability ($k_{rw}$).

### 5.1. Critical water saturation

In this study, the critical water saturation for percolation was estimated from continuum percolation theory (van der Marck, 1996; Rintoul, 2000), which is on the basis of randomly distributed overlapping void spheres in a solid matrix. Although we found $S_{wc} = 0.03$ well in agreement with the numerical simulations, the value of critical water saturation in porous media depends on both medium and transport properties e.g., pore space microstructure and geometry, topology, wettability, fluid characteristics, and system size.

Topology is quantified by parameters, such as pore coordination number, its distribution and average as well as the number of alternative pathways spanning pores within the medium and the number of isolated clusters (Arns et al., 2004). The important effect of the average pore coordination number was earlier addressed within bond percolation theory (Stauffer and Aharony, 1994; Sahimi, 2011; Hunt et al., 2014). Sok et al. (2002) demonstrated that topological properties e.g., the coordination number distribution had a substantial impact on the threshold in disordered and irregular networks. They concluded that a more complete description of network topology is required to predict critical saturation for percolation precisely.

The effect of wettability on critical saturation for oil, gas, and water flow has been widely documented (see e.g., Anderson, 1987a; Blunt, 2017). Recently, Ghanbarian et al. (2015b) investigated the critical saturation for diffusion in mono-sized sphere packs. By comparing diffusion in lattice-Boltzmann simulations with identical pore space characteristics, e.g., structure and pore size distribution, Ghanbarian et al. (2015b) found that the critical saturation under perfectly wetting conditions was less than that under neutrally (or intermediately) wetting conditions. This means more water was required to form a percolating pathway spanning the medium as the contact angle changed from perfectly wet to neutrally wet. Similarly, more gas content (larger critical saturation) was needed to form a continuous pathway for gas. It is generally expected that the critical saturation for nonwetting-phase flow to be slightly less than that for wetting-phase capillary flow because of capillary bridges (or pendular structures) and disconnected wetting films, which occupy volume without contributing to capillary flow (Ghanbarian et al., 2014). We, however, caution that if there exist films effectively contributing to the fluid flow, in parallel to capillary flow, one should expect a very small threshold at low saturation because thin films on pores surface allow the fluid to flow.

The system size dependence of the critical saturation has been also well demonstrated in the literature. For example, Wilkinson and Willemsen (1983) showed that the volume fraction for percolation scales with the system size (or length) in the power-law form. Yortsos and his coworkers (Li and Yortsos, 1995; Du and Yortsos, 1999) also found that critical saturation scaled with the network size and the fraction of network sites where gas nucleation takes places.

## 5.2. Water relative permeability

Within CPA framework, $k_{rw}$ is mainly controlled by the conductance or pore size distribution. In real rocks and sediments, however, $k_{rw}$ depends on others factors as well. For example, the effects of wettability, capillary number – a dimensionless quantity measuring the ratio of viscous to capillary forces – and trapping number – a dimensionless parameter quantifying the interaction of viscous, gravitational, and capillary forces – may be nontrivial (Pope et al., 2000; Beygi et al., 2015). Anderson (1987b) stated that, "Wettability affects relative permeability because it is a major factor in the control of the location, flow, and spatial distribution of fluids in the core." A historical account of the investigation of the wettability effect on $k_{rw}$ up to the 1980s was given by Anderson (1987b).

We should point out that the proposed CPA-based models (Eqs. 7b and 8b) can be applied to estimate wetting-phase relative permeability in homogeneously-wet porous media in which the wetting phase is restricted to smaller pores, whereas the non-wetting phase to larger pores. As stated earlier, the effect of wettability on $k_{rw}$ was indirectly incorporated in our model in which the water relative permeability is estimated from the pore size distribution derived from the capillary pressure curve. Such a derivation requires employing the Young-Laplace equation and the contact angle value. The latter was set equal to zero, in accord with the assumption used in the numerical simulations of Mawer et al. (2015) and Silin and Patzek (2009). In reality, however, water is typically contaminated, and the soil pore-solid interface is rough, unsmooth, sub-critically water repellent and mineralogically heterogeneous. Therefore, one should expect the contact angle to be greater than zero. For precise estimation of $k_{rw}$, the contact angle should be accurately characterized. Since its value may change from one soil/rock sample to another, one should directly determine the value of contact angle. Particularly, Andrew et al. (2014) measured the contact angle between immiscible fluids at the pore scale through X-ray microtomography in a mixed-wet rock. They observed a distribution of contact angles ranging from 35 to 55° in a supercritical $CO_2$-brine-carbonate system and found contact angle measured under the imbibition process greater than that determined under the drainage one.

The effect of capillary number was not incorporated into our CPA-based model. However, evidence reported in the literature (see e.g., Fulcher et al., 1985; Ostos and Maini, 2004; Li et al., 2005) indicate that as capillary number increases, $k_{rw}$ should increase as well. Trapping number should also have similar impact on $k_{rw}$ (see Fig. 2 of Pope et al., 2000). Hunt and Manga (2003) mapped a porous medium into a network of pore tubes and applied critical path analysis to study the effects of (air) bubble dynamics on water relative permeability. They found that water relative permeability increased monotonically with increasing capillary number (Ca) up to $10^{-2}$, but might decrease for higher capillary numbers due to the relative decrease of bubble density in the critical pores. Given that Hunt and Manga (2003) did not evaluate their model with experiments or simulations, further investigation is required to appropriately incorporate the effect of capillary number into the CPA-based $k_{rw}$ model and compare it via experimental observations and/or numerical simulations.

**5.3. Numerical simulation of fluid flow**

One of the main challenges in the simulation of multiphase flow in porous media is that transport may involve the simultaneous coupling of viscous and capillary forces. There exist many approaches that demonstrate the decoupling of viscous and capillary forces into a serial mode such as quasi-static pore network modelling. The main justification is that one can potentially decompose the relative permeability problem into two steps: (1) computing the pore-scale fluid distribution and (2) computing the hydraulic conductivities (i.e., the relative permeabilities). The assumption is that fluid distributions can be computed based on capillary forces. That is, however, only the case for drainage processes in strongly hydrophilic porous media and at low capillary numbers (see e.g., Berg et al., 2016). In almost all other situations this assumption has been proven to be invalid. The second step, i.e. computing relative permeability only on a quasi-static basis, is also only valid for strongly water-wet situations and only for saturation ranges sufficiently far away from the percolation threshold (Armstrong et al., 2016). For mixed- and intermediate-wet situations, relevant to many practical applications, there is a substantial degree of ganglion dynamics which also transports significant flux (Rücker et al., 2019). Accordingly, it matters whether a simulation technique uses the full visco-capillary balance such as the lattice-Boltzmann method with moving liquid-liquid interfaces or whether a quasi-static approach such as the Al-Futaisi and Patzek (2003) method is used. The latter has been validated against synchrotron beamline data (Berg et al., 2016) and found to not work for the imbibition process.

## 6. Conclusions

In this study, two approaches were presented based on critical path analysis (CPA) to model and estimate water relative permeability $k_{rw}$ in mono-sized sphere packs, known as homogeneous porous media with relatively narrow pore size distributions. The main objectives were: (1) applying CPA to estimate water relative permeability $k_{rw}$ in packings of randomly distributed spheres of the same size, and (2) comparing $k_{rw}$ estimated from the pore size distribution with that predicted from both the pore size distribution and relative electrical conductivity. Using numerical simulations, we showed that although CPA was originally developed to model flow and transport in heterogeneous media with broad conductance or pore size distributions, the proposed models estimated $k_{rw}$ in mono-sized packings of spheres accurately. More specifically, we demonstrated that CPA estimated $k_{rw}$ from the pore size distribution and electrical conductivity data more precisely than merely the pore size distribution for both drainage and imbibition processes.

**Acknowledgment**

The author acknowledges the Associate Editor, WeiCheng Lo, and two anonymous reviewers for constructive and fruitful comments that improved the quality of this manuscript as well as Chloe Mawer, Silicon Valley Data Science, for providing the Mawer et al. (2015) database used in this study. The author is also grateful to Kansas State University for supports through faculty startup funds.

**Notation**

| | |
|---|---|
| $A$ | normalizing factor |
| $f(g)$ | conductance distribution |
| $f(r)$ | pore size distribution |
| $g$ | pore conductance |
| $g_c$ | critical pore conductance |
| $k_0$ | constant coefficient |
| $k_w$ | water permeability |
| $k_{rw}$ | water relative permeability |
| $P$ | capillary pressure |
| $P_m$ | capillary pressure corresponding to $r_m$ |
| $P_{min}$ | minimum capillary pressure |
| $P_{max}$ | maximum capillary pressure |
| $r$ | pore radius |
| $r_c$ | critical pore radius |
| $r_m$ | geometric mean pore radius |
| $r_{min}$ | minimum pore radius |
| $r_{max}$ | maximum pore radius |
| $s$ | shape factor |
| $S_e$ | effective water saturation |
| $S_w$ | water saturation |
| $S_{wc}$ | critical water saturation |
| $S_{wr}$ | residual water saturation |
| $S_{wx}$ | crossover water saturation |
| $t$ | universal scaling exponent |
| $\alpha$ | exponent in Poiseuille's law |
| $\phi$ | porosity |
| $\mu$ | non-universal scaling exponent |
| $\sigma$ | log-normal standard deviation |
| $\sigma_b$ | bulk electrical conductivity |
| $\sigma_r$ | relative electrical conductivity |

Fire Research Laboratory, National Institute of Standards and Technology. Springfield, VA.

Ghanbarian, B. (2018). Estimating gas relative permeability of shales from pore size distribution. In SPE Annual Technical Conference and Exhibition. Society of Petroleum Engineers. Dallas TX. SPE-191878-MS.

Ghanbarian, B. (2019). Applications of critical path analysis to uniform grain packs with narrow local conductance distributions: I. Single-phase permeability. Water Resources Research (under review).

Ghanbarian-Alavijeh, B., and A. G. Hunt (2012a), Unsaturated hydraulic conductivity in porous media: Percolation theory. Geoderma, 187, 77-84.

Ghanbarian-Alavijeh, B., Hunt, A. G. (2012b). Comparison of the predictions of universal scaling of the saturation dependence of the air permeability with experiment. Water Resources Research, 48(8), W08513.

Ghanbarian, B., and Hunt, A. G. (2017). Improving unsaturated hydraulic conductivity estimation in soils via percolation theory. Geoderma, 303, 9-18.

Ghanbarian, B., & Sahimi, M. (2017). Electrical conductivity of partially saturated packings of particles. Transport in Porous Media, 118(1), 1-16.

Ghanbarian, B., Hunt, A. G., Ewing, R. P., and Skinner, T. E. (2014), Theoretical relationship between saturated hydraulic conductivity and air permeability under dry conditions: Continuum percolation theory, Vadose Zone Journal, 13(8).

Ghanbarian, B., Hunt, A. G., Skinner, T. E., and Ewing, R. P. (2015a). Saturation dependence of transport in porous media predicted by percolation and effective medium theories. Fractals, 23(01), 1540004.
32

Jerauld, G. R., and S. J. Salter (1990), The effect of pore-structure on hysteresis in relative permeability and capillary pressure: pore-level modeling, Transport in Porous Media, 5(2), 103-151.

Katz, A. J., and A. H. Thompson (1986). Quantitative prediction of permeability in porous rock. Physical review B, 34(11), 8179-8181.

Kosugi, K. I. (1994). Three-parameter lognormal distribution model for soil water retention. Water Resources Research, 30(4), 891-901.

Kosugi, K. I. (1996). Lognormal distribution model for unsaturated soil hydraulic properties. Water Resources Research, 32(9), 2697-2703.

Landry, C. J., Z. T. Karpyn, and O. Ayala (2014), Relative permeability of homogenous-wet and mixed-wet porous media as determined by pore- scale lattice Boltzmann modeling, Water Resour. Res., 50, 3672-3689.

Larson, R. G., L. E. Scriven, and H. T. Davis (1981), Percolation theory of two phase flow in porous media, Chemical Engineering Science, 36(1), 57-73.

Li, X., and Yortsos, Y. C. (1995). Theory of multiple bubble growth in porous media by solute diffusion. Chemical Engineering Science, 50(8), 1247-1271.

Li, H., C. Pan, and C. T. Miller (2005), Pore-scale investigation of viscous coupling effects for two-phase flow in porous media, Phys. Rev. E, 72, 026705.

Madadi, M., and Sahimi, M. (2003). Lattice Boltzmann simulation of fluid flow in fracture networks with rough, self-affine surfaces. Physical Review E, 67(2), 026309.

Mawer, C., R. Knight, and P. K. Kitanidis (2015), Relating relative hydraulic and electrical conductivity in the unsaturated zone, Water Resources Research, 51(1), 599-618.
35

Table 1. The log-normal probability density function parameters for 15 packs from Mawer et al. (2015).

| Pack | Porosity | Condition | A | $\sigma$ | $r_m$ (μm) | $r_{min}$ (μm) | $r_{max}$ (μm) | $R_2$ |
|---|---|---|---|---|---|---|---|---|
| Finney | 0.362 | Drainage | 281.7 | 0.212 | 328.3 | 0 | 997.2 | 0.97 |
|  |  | Imbibition | 368.4 | 0.351 | 444.6 | 0 | 1200.0 | 0.96 |
| 1 | 0.335 | Drainage | 7.8 | 0.218 | 9.4 | 0 | 27.5 | 0.96 |
|  |  | Imbibition | 40.8 | 0.361 | 49.9 | 0 | 160.0 | 0.94 |
| 2 | 0.351 | Drainage | 8.1 | 0.190 | 9.7 | 0 | 23.6 | 0.97 |
|  |  | Imbibition | 113.5 | 0.362 | 135.1 | 0 | 350.0 | 0.92 |
| 3 | 0.230 | Drainage | 5.8 | 0.272 | 6.6 | 0 | 13.8 | 0.93 |
|  |  | Imbibition | 35.8 | 0.429 | 41.1 | 0 | 150.0 | 0.88 |
| 4 | 0.249 | Drainage | 5.8 | 0.279 | 6.6 | 0 | 17.7 | 0.96 |
|  |  | Imbibition | 32.7 | 0.380 | 39.6 | 0 | 120.0 | 0.92 |
| 5 | 0.261 | Drainage | 5.6 | 0.179 | 7.2 | 0 | 15.7 | 0.96 |
|  |  | Imbibition | 35.8 | 0.402 | 41.7 | 0 | 120.0 | 0.90 |
| 6 | 0.280 | Drainage | 6.0 | 0.198 | 7.4 | 0 | 19.7 | 0.98 |
|  |  | Imbibition | 35.3 | 0.374 | 42.5 | 0 | 120.0 | 0.93 |
| 7 | 0.300 | Drainage | 5.9 | 0.184 | 7.3 | 0 | 15.7 | 0.98 |
|  |  | Imbibition | 39.7 | 0.398 | 45.0 | 0 | 120.0 | 0.92 |
| 8 | 0.317 | Drainage | 6.6 | 0.216 | 7.8 | 0 | 17.7 | 0.99 |
|  |  | Imbibition | 40.4 | 0.400 | 46.5 | 0 | 150.0 | 0.92 |
| 9 | 0.337 | Drainage | 7.0 | 0.214 | 8.2 | 0 | 21.6 | 0.97 |
|  |  | Imbibition | 38.1 | 0.351 | 45.6 | 0 | 120.0 | 0.95 |
| 10 | 0.358 | Drainage | 7.4 | 0.204 | 8.7 | 0 | 19.7 | 0.96 |
|  |  | Imbibition | 41.5 | 0.362 | 48.7 | 0 | 120.0 | 0.93 |
| 11 | 0.376 | Drainage | 7.7 | 0.205 | 9.0 | 0 | 17.7 | 0.97 |
|  |  | Imbibition | 44.9 | 0.369 | 50.6 | 0 | 150.0 | 0.94 |
| 12 | 0.419 | Drainage | 8.4 | 0.189 | 9.9 | 0 | 21.6 | 0.97 |
|  |  | Imbibition | 46.5 | 0.341 | 53.8 | 0 | 150.0 | 0.92 |
| 13 | 0.438 | Drainage | 8.7 | 0.178 | 10.3 | 0 | 21.6 | 0.97 |
|  |  | Imbibition | 48.9 | 0.342 | 55.4 | 0 | 150.0 | 0.94 |
| 14 | 0.458 | Drainage | 9.3 | 0.156 | 10.8 | 0 | 23.6 | 0.97 |
|  |  | Imbibition | 49.4 | 0.315 | 57.0 | 0 | 150.0 | 0.93 |
| Ave. | 0.338 (0.065)* | Drainage | 25.5 (70.9) | 0.206 (0.03) | 29.8 (82.6) | 0 (0) | 85.0 (252.4) | 0.97 (0.01) |
|  |  | Imbibition | 67.4 (85.5) | 0.369 (0.03) | 79.8 (103.5) | 0 (0) | 222.0 (276.5) | 0.93 (0.02) |

* Numbers in parentheses represent standard deviations.

**Figure captions**

Figure 1. Two-dimensional scheme of the critical path analysis. (a) A pore network compsed of six different pore sizes (i.e., 0.5, 1, 1.5, 2, 2.5 and 3 with arbitrary units) randomly distributed in the medium. (b) The same network with only the first two largest pores (2.5 and 3) in their original locations. Pores smaller than 2.5 were removed from the pore network. As can be seen, the medium does not percolate. (c) The nerwork after adding the third largest pores with size 2 (critical pore size). The sample-spanning cluster is first formed and the network starts percolating.

Figure 2. Pore size distributions, derived from simulated *drainage* capillary pressure curves, for 15 mono-sized sphere packs from Mawer et al. (2015). The red line represents the fitted log-normal distribution, Eq. (1). Table 1 summarizes the optimized parameters.

Figure 3. Water relative permeability $k_{rw}$ as a function of water saturation $S_w$ during *drainage* for 15 mono-sized sphere packs from Mawer et al. (2015). The blue and black lines represent the critical-path-analysis scaling, while the red and grey lines denote the universal power-law scaling from percolation theory. The $k_{rw}$ curve estimated from the PSD is represented by black and gray lines, while that predicted from the PSD and $\sigma_r$ is denoted by blue and red lines.

Figure 4. (a) Simulated capillary pressure curves, (b) derived pore size distributions, and (c) simulated water relative permeability curves during *drainage* for 15 mono-sized sphere packs from Mawer et al. (2015). (d) Logarithm of simulated $k_{rw}$ as a function of water saturation. The red line denotes the corresponding curve in each pack. The black

line represents the Kozeny-Carman equation adopted for partially-saturated conditions (i.e., $k_{rw} = S_w^3$).

Figure 5. Pore size distributions, derived from simulated *imbibition* capillary pressure curves, for 15 mono-sized sphere packs from Mawer et al. (2015). The red line represents the fitted log-normal distribution, Eq. (1). Table 1 summarizes the optimized parameters.

Figure 6. Water relative permeability $k_{rw}$ as a function of water saturation $S_w$ during *imbibition* for 15 mono-sized sphere packs from Mawer et al. (2015). The blue and black lines represent the critical-path-analysis scaling, while the red and grey lines denote the universal power-law scaling from percolation theory. The $k_{rw}$ curve estimated from the PSD is represented by black and gray lines, while that predicted from the PSD and $\sigma_r$ is denoted by blue and red lines.

Figure 7. (a) Simulated capillary pressure curves, (b) derived pore size distributions, and (c) simulated water relative permeability curves during *imbibition* for 15 mono-sized sphere packs from Mawer et al. (2015). (d) Logarithm of simulated $k_{rw}$ as a function of water saturation. The blue line denotes the corresponding curve in each pack. The black line represents the Kozeny-Carman equation adopted for partially-saturated conditions (i.e., $k_{rw} = S_w^3$).

Figure 8. Water relative permeability $k_{rw}$ as a function of water saturation $S_w$ during *drainage* for a Finney pack from Silin and Patzek (2009). The blue and black lines represent the critical-path-analysis scaling, while the red and grey lines denote the universal power-law scaling from percolation theory. The log-normal pore size distribution parameters and porosity were estimated from Finney pack given in Table 1.

We set $S_{wr} = 0$ and $S_{wc} = 0.03$. The $k_{rw}$ curve estimated from the PSD is represented by black and gray lines, while that predicted from the PSD and $\sigma_r$ is denoted by blue and red lines.

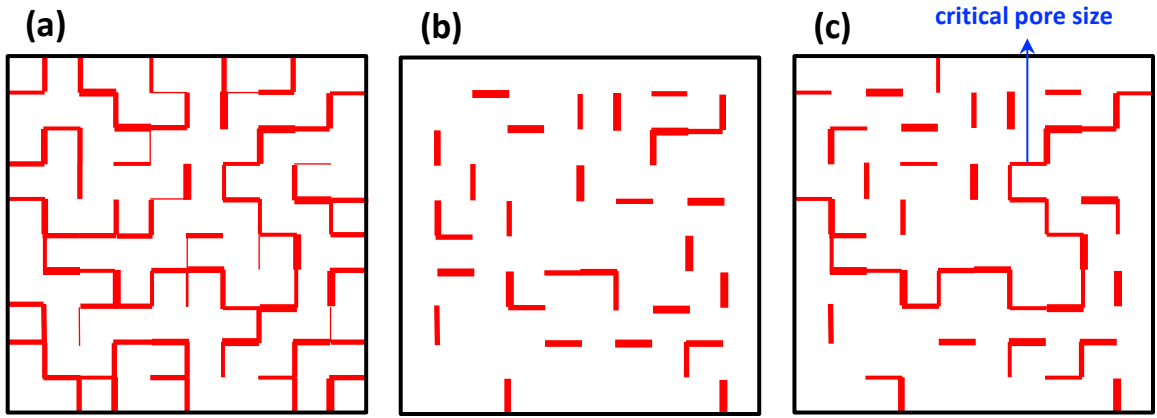

Fig. 1

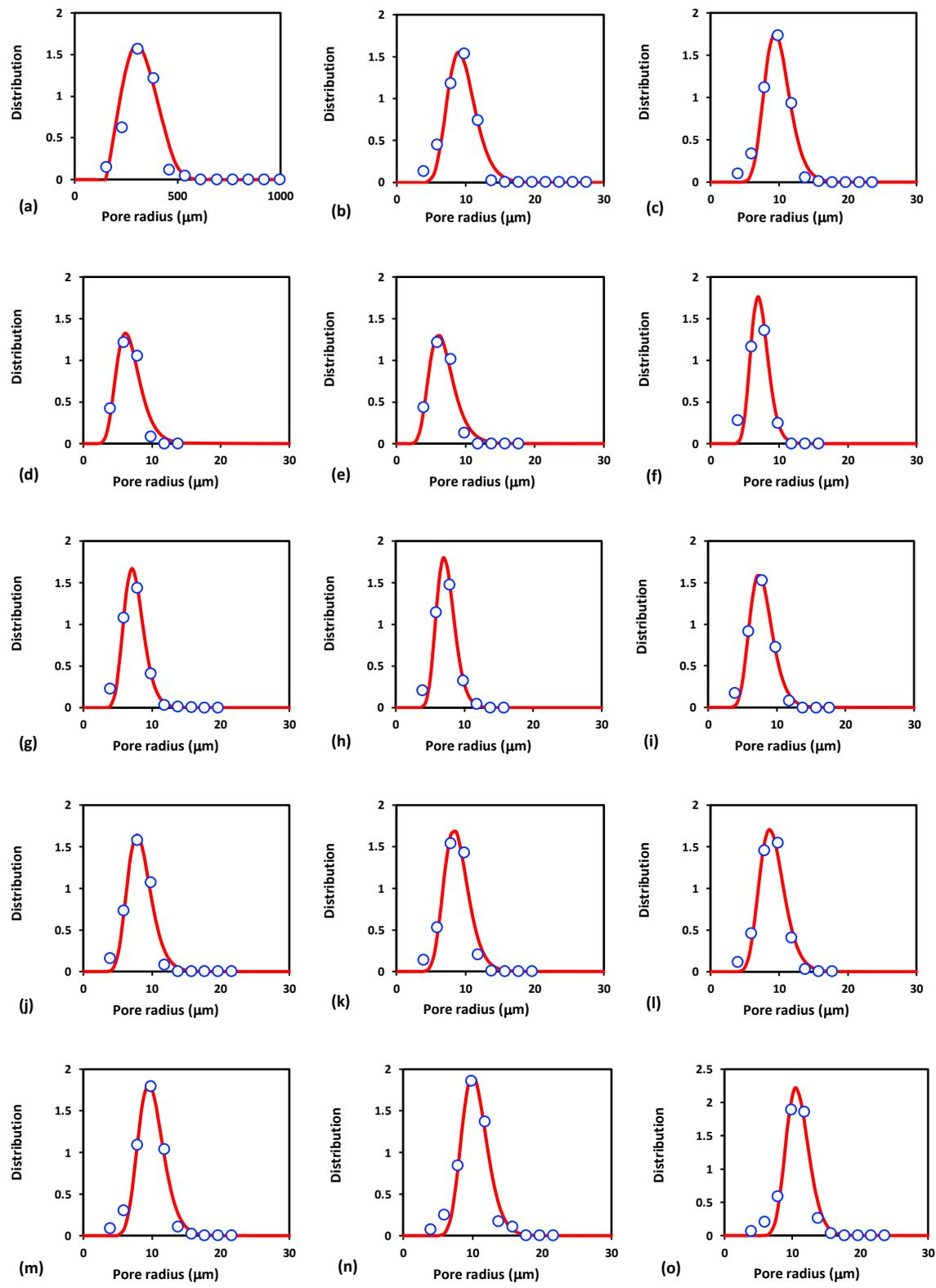

Fig. 2

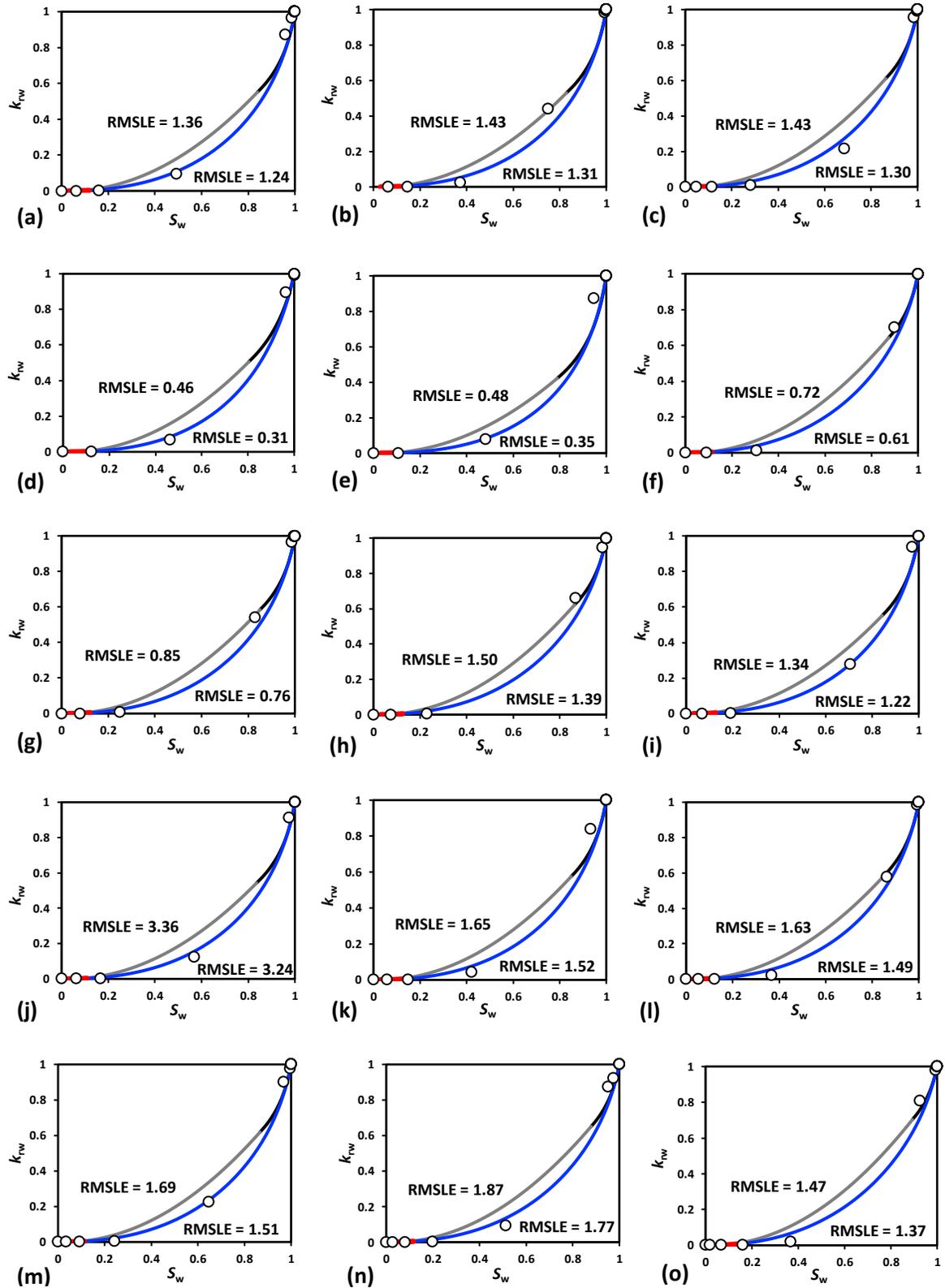

Fig. 3

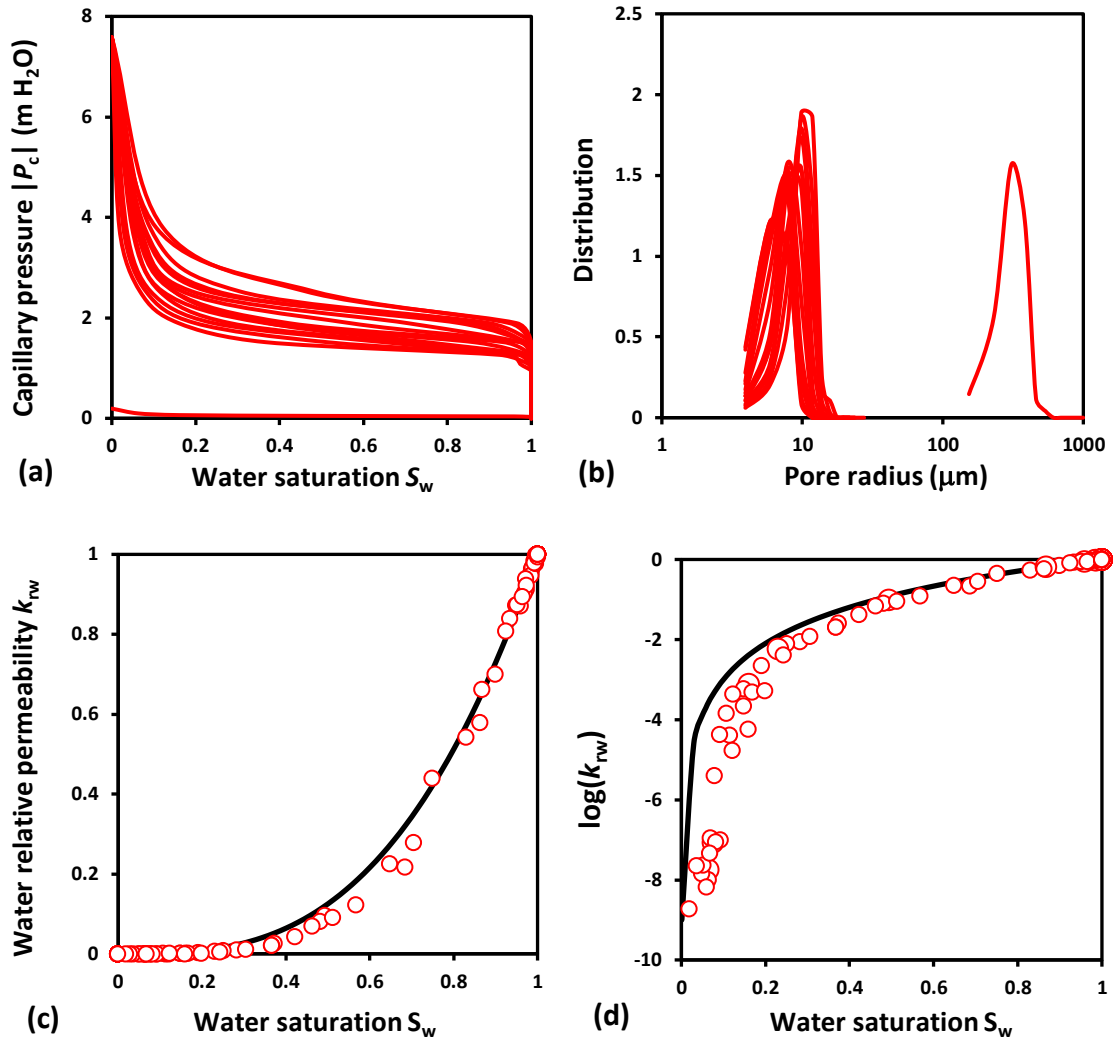

Fig. 4

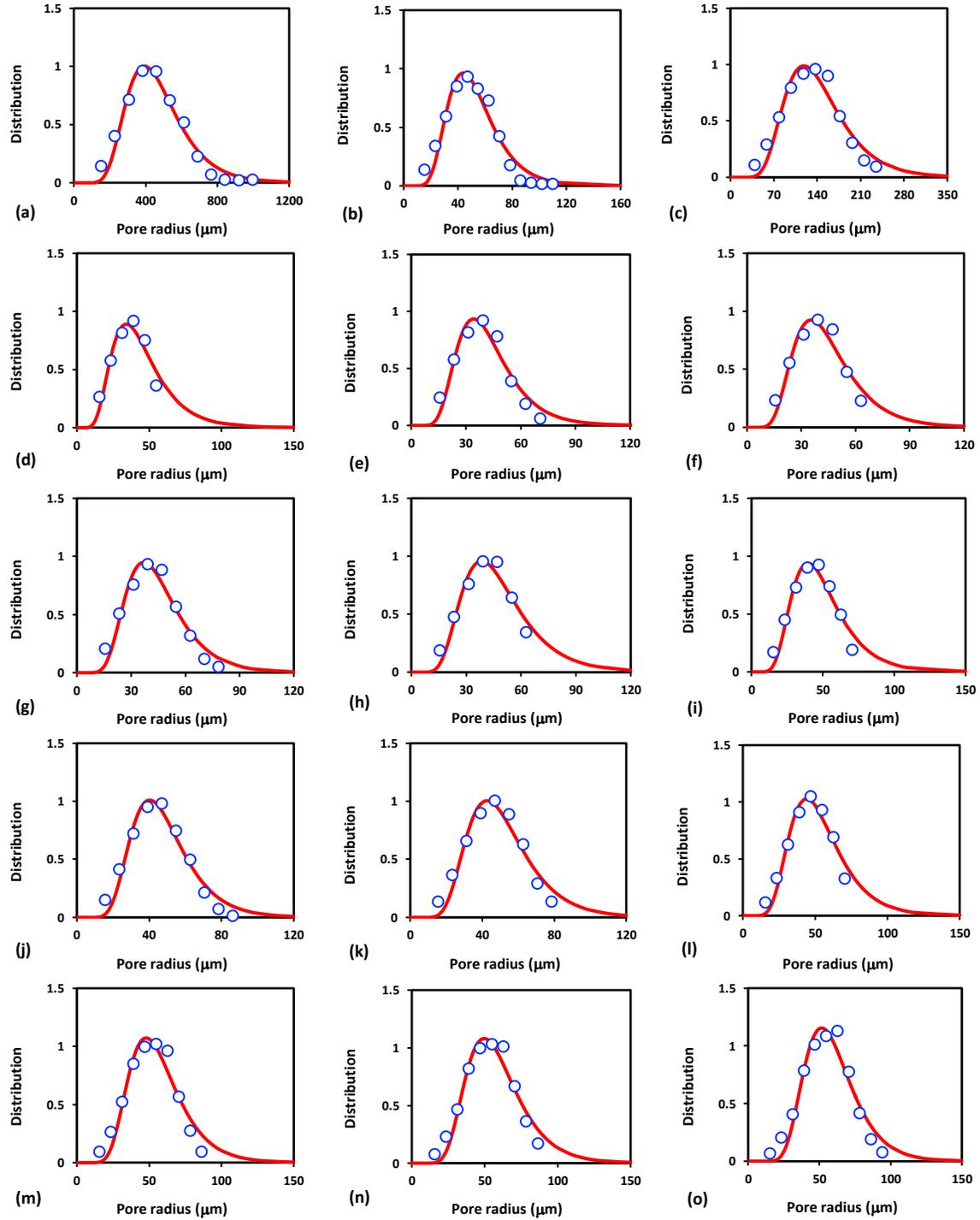

Fig. 5

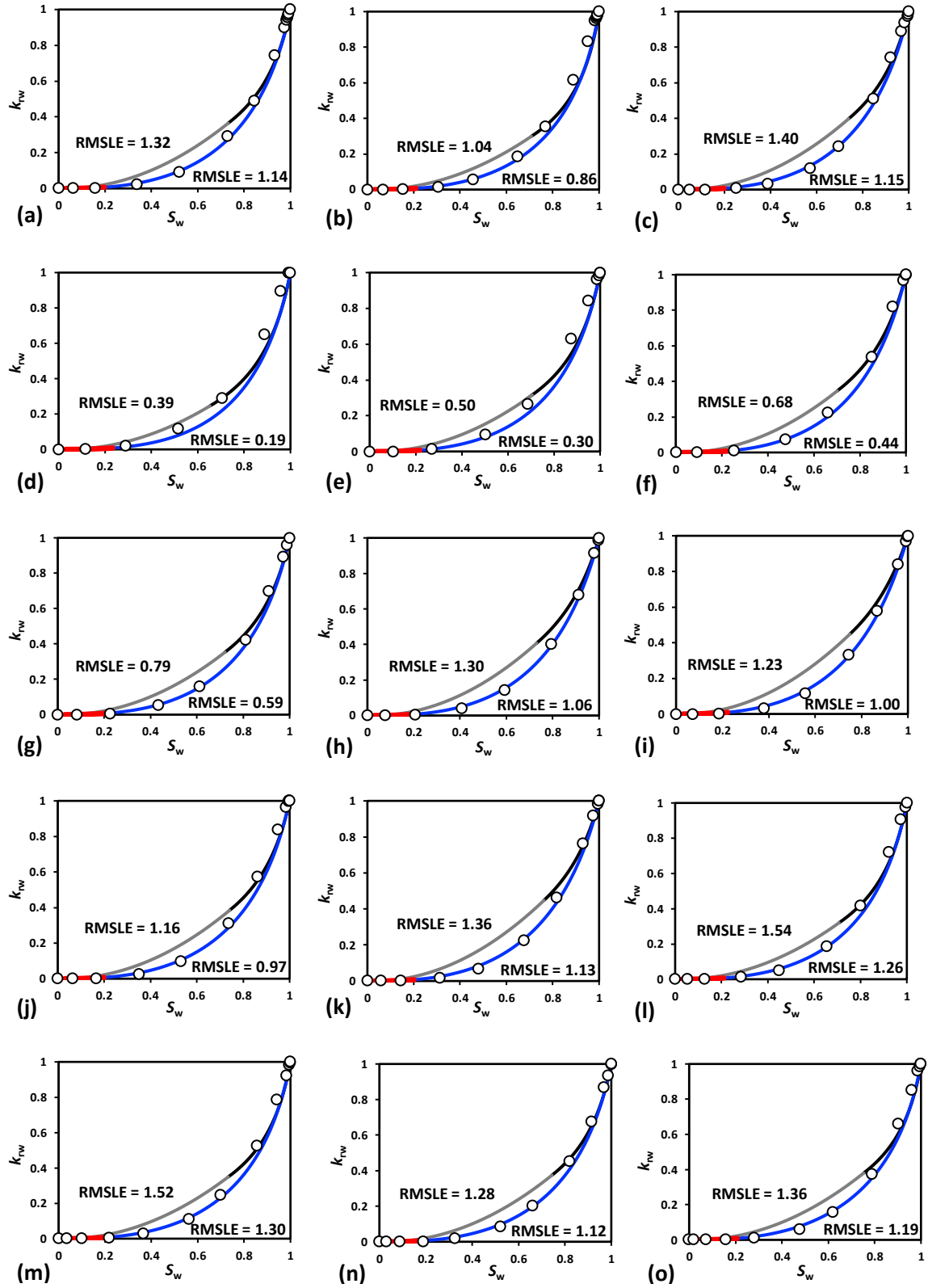

Fig. 6

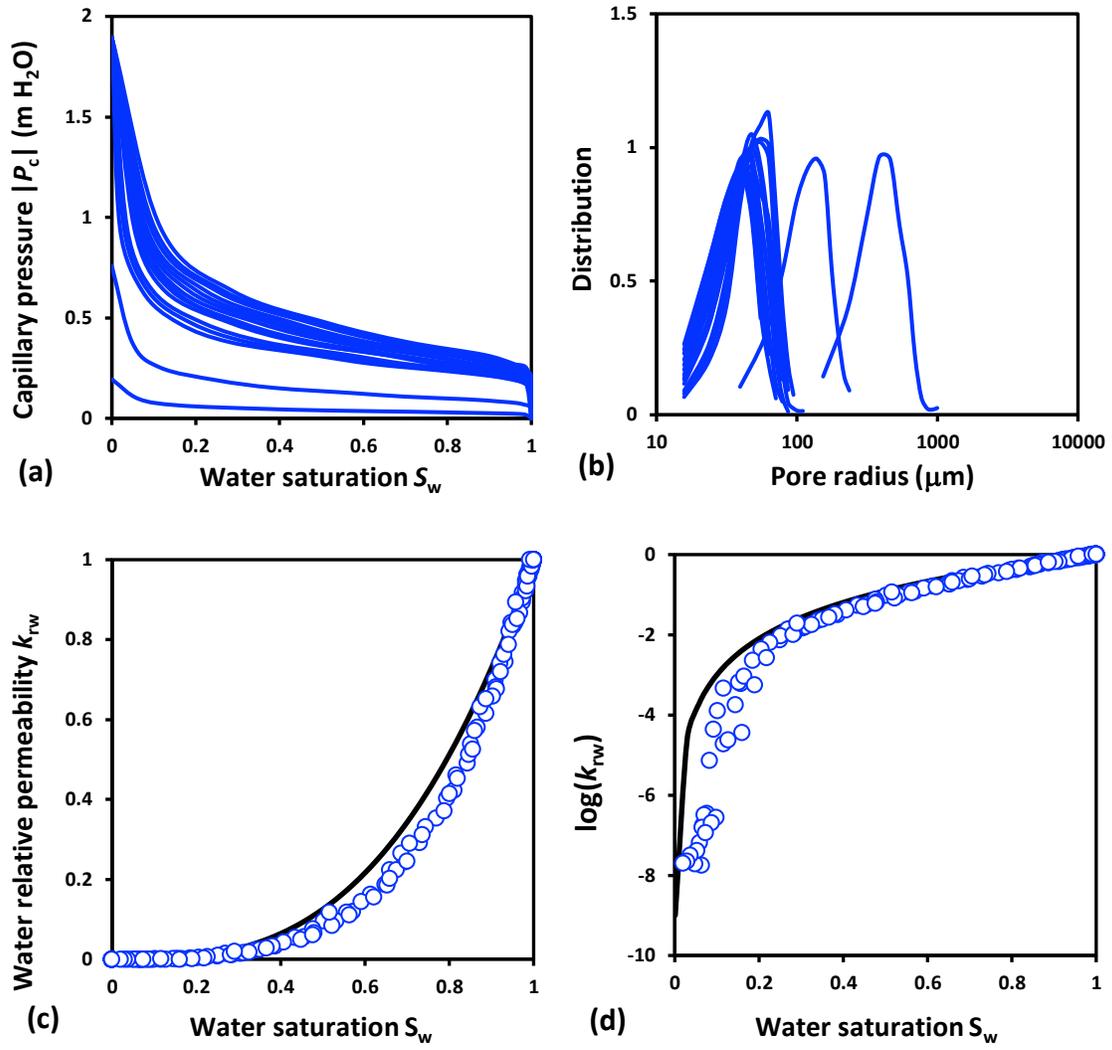

Fig. 7

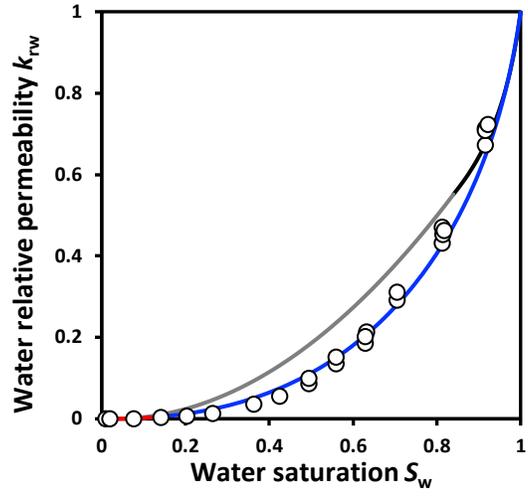

Fig. 8